# Tunable lattice reconstruction and bandwidth of flat bands in magic-angle twisted bilayer graphene

Yi-Wen Liu[1,§], Ying Su[2,§], Xiao-Feng Zhou[1], Long-Jing Yin[1,3], Chao Yan[1], Si-Yu Li[1,4], Wei Yan[1], Sheng Han[1], Zhong-Qiu Fu[1], Yu Zhang[1], Qian Yang[1], Ya-Ning Ren[1], and Lin He[1,†]

[1] Center for Advanced Quantum Studies, Department of Physics, Beijing Normal University, Beijing 100875, People's Republic of China

[2] Theoretical Division, T-4 and CNLS, Los Alamos National Laboratory, Los Alamos, New Mexico 87545, USA

[3] Key Laboratory for Micro/Nano Optoelectronic Devices of Ministry of Education & Hunan Provincial Key Laboratory of Low-Dimensional Structural Physics and Devices, School of Physics and Electronics, Hunan University, Changsha 410082, China.

[4] Key Laboratory for Micro-Nano Physics and Technology of Hunan Province, College of Materials Science and Engineering, Hunan University, Changsha 410082, People's Republic of China

[§]These authors contributed equally to this work.

[†]Correspondence and requests for materials should be addressed to L.H. (e-mail: helin@bnu.edu.cn).

**The interplay between interlayer van der Waals interaction and intralayer lattice distortion can lead to structural reconstruction in slightly twisted bilayer graphene (TBG) with the twist angle being smaller than a characteristic angle $\theta_c$. Experimentally, the $\theta_c$ is demonstrated to be very close to the magic angle ($\theta \approx 1.05°$). In this work, we address the transition between reconstructed and unreconstructed structures of the TBG across the magic angle by using scanning tunnelling microscopy (STM). Our experiment demonstrates that both the two structures are stable in the TBG around the magic angle. By applying a STM tip pulse, we show that the two structures can be switched to each other and the bandwidth of the flat bands, which plays a vital role in the emergent strongly correlated states in the magic-angle TBG, can be tuned. The observed tunable lattice reconstruction and bandwidth of the flat bands provide an extra control knob to manipulate the exotic electronic states of the TBG near the magic angle.**

Twisted bilayer graphene (TBG) formed by a vertical stacking of two misaligned graphene layers has increasingly attracted considerable interests due to the efficient tunability of electronic band structures by changing twist angles[1-7]. Remarkably, at the so-called magic angle ($\theta \approx 1.05°$), the low-energy bands around the charge neutrality become nearly flat. As a consequence, many strongly correlated states are observed in the TBG because the substantial quench of the kinetic energy of quasiparticles, which is determined by electronic bandwidth of the flat bands[8-13]. Very recently, it was demonstrated that there is structural reconstruction in slightly TBG because the competition between the interlayer van der Waals coupling and the intralayer elastic deformation[14-21]. In the reconstructed TBG, the energetically preferable AB/BA stacking regions are enlarged, while the AA stacking regions are shrunken and corrugated to reduce the cost of energy. Experimentally[19], the structural reconstruction occurs in the TBG with twist angle $\theta \leq \theta_c \sim 1°$. Obviously, the characteristic crossover angle $\theta_c$ between the reconstructed and unreconstructed structures is very close to the magic angle.

In this work, we systematically studied the structures of the TBG around the magic angle by using scanning tunnelling microscopy (STM) and demonstrated that both the reconstructed and unreconstructed structures can be stable around the magic angle. By applying a STM tip pulse, the two structures in the TBG near the magic angle can be switched to each other. As a consequence, the bandwidth of the flat bands is changed dramatically. Our result not only offers a reasonable explanation to understand the quite different electronic properties observed in the magic-angle TBG, but also provides a new route to manipulate the exotic electronic phases of the magic-angle TBG.

In our experiments, monolayer graphene was synthesized by traditional low-pressure chemical vapor deposition (LPCVD) method on Cu foil obtained by contact-free annealing[22] (see Supplementary Fig. S1 for details of the growth). Large aligned monolayer graphene grown on the Cu foils is obtained (see Supplementary Fig. S2)[23-25]. Then, through conventional wet etching technique, the monolayer graphene was transferred layer-by-layer onto different target substrates, involving 0.7% Nb-doped SrTiO$_3$ (001), Si/SiO$_2$ wafer, Ag-coated mica, and Cu (111) substrates, to obtain the TBG with controlled twist angle (see Fig. S3 for details of the fabrication and Fig. S4 for STM characterizations of the substrates and the obtained TBG). Our experiment demonstrated a universal approach to fabricate layered van der Waals heterostructures with controlled twist angle on different substrates.

Despite the angle dependent electronic properties of the TBG has been extensively studied, the angle dependent lattice reconstruction is rarely explored in the STM studies. Figures 1a-1d show representative STM images of the obtained TBG with four different twist angles. The moiré superlattice can be clearly identified from the periodic corrugations at the AA stacking regions. For the TBG with $\theta = 17.76°$, the variation of height between $h_{AA}$ at the AA regions and $h_{AB/BA}$ at the AB/BA regions is only about 2 pm, as shown in Fig. 1e. However, it increases to about 60 pm for the TBG with $\theta \leq 4.2°$. The less lattice corrugation at larger twist angle is due to the fact that the interlayer coupling in TBG falls off rapidly with increasing twist angle[4,7]. For the TBG with $\theta = 0.87°$, as shown in Fig. 1d, we can observe clear signature of lattice reconstruction in the TBG: the AA regions are reduced while the AB/BA regions are enlarged to minimize the global energy of the system[14-21]. In the meantime, the domain walls (DWs) separating adjacent AB and BA regions are corrugated (see Fig. 1e, the height of the DWs is defined as $h_{DW}$) and form a triangular network connecting different AA regions. The rearrangement of atomic registries in reconstructed TBG compared to that in unreconstructed TBG is schematically shown in Fig. 1f.

According to a recent transmission electron microscopy study[19], the transition from structural un-reconstruction to structural reconstruction occurs in the TBG at the characteristic crossover angle $\theta_c \sim 1°$. In our STM measurements, we can obtain the evolution of structural reconstruction with the twist angle in the TBG in two different ways, as summarized in Fig. 1g. The first one is based on the relative strain between the DW and AB/BA regions as a function of the twist angle. Here the relative strain is estimated from the fast Fourier transformation (FFT) of different regions in the STM topographic images, as shown in the insets of Fig. 1g (see Supplementary Fig. S5 for more details), and it increases quite quickly with decreasing the twist angle below the characteristic crossover angle. The second method is based on the height ratio between the DW and AA regions $h_{DW}/h_{AA}$ as a function of the twist angle (Fig. 1g), which shows a similar behavior as the relative strain *vs* the twist angle. Obviously, the characteristic crossover angle for the structural reconstruction in the TBG is around the magic angle. Below the magic angle, the in-plane atomic registries in the TBG are rearranged significantly to minimize the global energy and the reconstructed TBG is the stable structure. Above the magic angle, the TBG prefers to be in the almost rigid structure.

The structural reconstruction is expected to dramatically change the electronic properties of the TBG[18-21]. One of the most prominent features in the reconstructed TBG is the emergence of AB-BA DWs that form a triangular network connecting different AA regions. In this case, the AB and BA regions are gapped and have opposite valley Chern numbers that guarantee the appearance of topological helical edge states at the DWs[26-33]. Such a feature can be directly imaged in the STM measurements[32,33]. To clearly show this, we carried out the energy-fixed scanning tunneling spectroscope (STS) mapping, which reflects the local density of states (LDOS) in real space[32-40]. Figure 2 shows three representative STS maps of three TBGs with different twist angles. For the TBG with $\theta = 2.68°$, which is larger than the magic angle, the distribution of the LDOS reveals the same period and symmetry of the moiré pattern in the STM image (see Supplementary Fig. S6 for STM image and STS spectra). Obviously, no signal of the DW can be observed in the 2.68 °TBG (Fig. 2a). As the twist angle decreases to $\theta = 1.08°$, which is around the magic angle, the triangular network of the AB-BA DWs connecting the AA regions appears, as shown in Fig. 2b. Additionally, bright halos around the dark AA regions, as observed in previous studies[32,41], are also observed in the 1.08 °TBG. With further lowering the twist angle to $\theta = 0.27°$, the conspicuous topological channels linking different AA stacking regions can be more clearly identified in Fig. 2c. Then, the width of the DW can be as large as 10 nm, which allows us to observe the characteristic feature that the topological states are mainly located at the two edges of the DW, as observed previously in wide DWs[28,29].

In the TBG, the structural reconstruction arises from the competition between the interlayer van der Waals coupling and the intralayer elastic deformation[14-21]. For the TBG with the twist angle below the magic angle, the size of the moiré pattern is quite large and the interlayer coupling is relatively strong. Therefore, the AB/BA stacking regions prefer to be enlarged to reduce the cost of stacking energy and, consequently, the TBG is reconstructed. For the TBG with the twist angle that is much larger than the magic angle, the size of the moiré pattern is quite small and the interlayer coupling is very weak. Hence the rigid TBG structure can minimize the intralayer elastic energy and is energetically preferable. Therefore, in the crossover region, *i.e.*, the TBG with the twist angle around the magic angle, the global energy for the reconstructed and un-reconstructed structures should be comparable. In Fig. 3a, we show the schematic energy profiles in the configuration space that illustrate the structural transition from unreconstructed TBG to reconstructed TBG as the twist angle decreases across the magic angle (or the $\theta_c$). Around the magic angle (or the $\theta_c$), the bi-stable structures are separated by an energy barrier and it is expected to observe the two structures simultaneously. However, direct observation of the two structures in the TBG at the crossover angle is still lacking up to now. In our experiment, we directly observed both the reconstructed and un-reconstructed structures in the TBG around the magic angle. Figures 3b-3e summarize STM measurements on a TBG with $\theta = 1.15°$. Both the STM topographic image and STS map of the 1.15 °TGB show the coexistence of the two structures that are separated by a boundary, as marked by the dashed line in Fig. 3b and 3d. Here we denote the area above (below) the boundary as region I (II). Even though the periods of the moiré pattern (or the twist angles) in the two regions are the same, much clearer DW structures are observed in the region II, as shown in Fig. 3b-3e. The topological edge states along the AB-BA DWs that form a triangular network connecting different AA regions are clearly observed in the region II. However, the topological network is absent in the region I. Such a result demonstrates explicitly that the region I of the 1.15 °TGB is in the un-reconstructed structure and the region II of the 1.15 °TGB is in the reconstructed structure.

Since both the reconstructed and un-reconstructed structures can be stable in the TBG around the magic angle, therefore, it is reasonable to ask whether the two structures can be switchable. Previously, it has been demonstrated that the STM tip can reshape the topography of underneath graphene via the electrostatic force and/or van der Waals force[42-50]. In this work, we demonstrated that the two structures in the TBG near the magic angle can be switched to each other by applying a STM tip pulse. Figure 4 summarizes the result obtained in a 1.13 °TBG. Before applying a STM tip pulse, the 1.13 °TBG is in the reconstructed structure and exhibits sharp DWs linking different AA regions in the STM topographic image, as shown in Fig. 4a. After applying a 3-V tip pulse for 0.1 s, the morphology of the TBG changes a lot under the same imaging bias. The sharp DW network almost vanishes and the size of AA stacking region enlarges, as shown in Fig. 4b (see Supplementary Fig. S7 for FFT images of two cases). Figure 4c shows the height profiles along the dashed arrows in Fig. 4a and 4b. Six well-defined corrugated DWs can be detected around each AA region before applying the tip voltage and they become less distinct after the pulse. Here the electrostatic force may play a dominant role because of the short time and large value of the tip voltage. This sudden applied energy perturbation leads to a structural transition of the TBG. In our experiment, we also find that the structure of the TGB can be tuned back to its original state by the tip pulse (see Supplementary Fig. S8). The change of lattice reconstruction before and after applying the tip voltage suggests that the STM tip not only can image the structure and measure the LDOS at nanoscale, but also can reshape the topography of the TBG effectively via a pulse of tunneling voltage, which overcomes the energy barrier between the bi-stable structures.

The change of lattice reconstruction in the 1.13 °TBG strongly alters its electric band structure, which can be directly measured by the STS spectra, as shown in Fig. 4d and 4e. The spectra recorded in the AA regions show low-energy sharp peaks, which are the flat bands of the TBG[33,34,38-40,51]. We can make three observations from the spectra. First, the doping of the TBG is changed: the flat bands are changed from fully empty (Fig. 4d) to fully occupied (Fig. 4e), which may partially arise from the variation of charge transfer between the TBG and the substrate because of the stimulus bias of the STM tip. Additionally, the enhanced next-nearest-neighbor hopping in strained TBG also can change the electron doping[33]. Second, the total full width at half maximum (FWHM) of the three flat bands in the reconstructed TBG (Fig. 4d), ~105 meV, is larger than that in the un-reconstructed TBG (Fig. 4e), ~70 meV. The third feature is the appearance of negative differential conductance (NDC) between the flat bands and the high-energy conduction band in the spectrum recorded in the reconstructed TBG (Fig. 4d), which is a clear signature that there is a gap between the flat bands and the remote bands[52,53] (see Supplementary Fig. S9 for another sample around the magic angle). To further understand the observed phenomena, we calculated the band structure and DOS of the magic-angle TBG with and without lattice relaxation (see Supplementary Fig. S10 for details). The essential features of the broadened and isolated flat bands in the reconstructed TBG can be reproduced by our theoretical calculation. However, the extra splitting (three subpeaks in Fig. 4d and 4e) of the flat bands may be induced by the strain effects in the TGB according to the previous studies[53-55], which is not involved in the calculation. Very recently, it was demonstrated explicitly that the heterostrain can efficiently reconstruct the band structure and lead to three flat bands in the TBG around the magic angle[54,55]. Therefore, the visible on and off DW network, the enlarged AA stacking regions, and the variation of corresponding band structures, as observed in this work, demonstrate explicitly that the lattice reconstruction in the TBG around the magic angle can be manipulated locally via STM tip pulses. Because the electronic correlation in the magic-angle TBG depends on the ratio between Coulomb interaction and bandwidth of the flat bands[8-13,38-40,51], therefore, our result provides a new route to manipulate the correlated electronic phases.

Similar measurement, i.e., by applying a pulse of tunneling voltage, has been carried out in the TBG with twist angle that is much larger (or smaller) that the magic angle in our experiment (see Supplementary Fig. S11 and Fig. S12). However, structures of the TBG below and above the magic angle are quite stable and almost the same before and after the tip pulse.

In summary, we report the tunable lattice reconstruction and bandwidth of the flat bands in the TBG around the magic angle. Our experiment demonstrates that both the reconstructed and un-reconstructed structures are stable in the TBG around the magic angle. The bi-stable structures are separated by an energy barrier and can be switched to each other by using a STM tip pulse that overcomes the energy barrier. Since the electronic correlation is determined by the flat band width, our experiments provide a method to manipulate the electronic property of the magic-angle TBG through the tunable lattice reconstruction. Such a result can be applied in transition metal dichalcogenides twisted bilayers[56] to tune their structures and electronic properties.

**Acknowledgements**

This work was supported by the National Natural Science Foundation of China (Grants No. 11974050 and No. 11674029). L. H. also acknowledges support from the National Program for Support of Top-notch Young Professionals, support from "the Fundamental Research Funds for the Central Universities," and support from "Chang Jiang Scholars Program." Y. S. was supported by the U.S. Department of Energy through the Los Alamos National Laboratory LDRD program, and was supported by the Center for Non-linear Studies at LANL.

**Author contributions**

Y.W.L., L.J.Y., C.Y., S.Y.L., W.Y., S.H., and Z.Q.F. performed the STM experiments. Y.W.L. and X.F.Z synthesized the samples and data. Y.S. performed the theoretical calculations. L.H. conceived and provided advice on the experiment, analysis, and the theoretical calculation. L.H., Y.W.L and Y.S. wrote the paper. All authors participated in the data discussion.

**Competing financial interests**

The authors declare no competing financial interests.

Figures

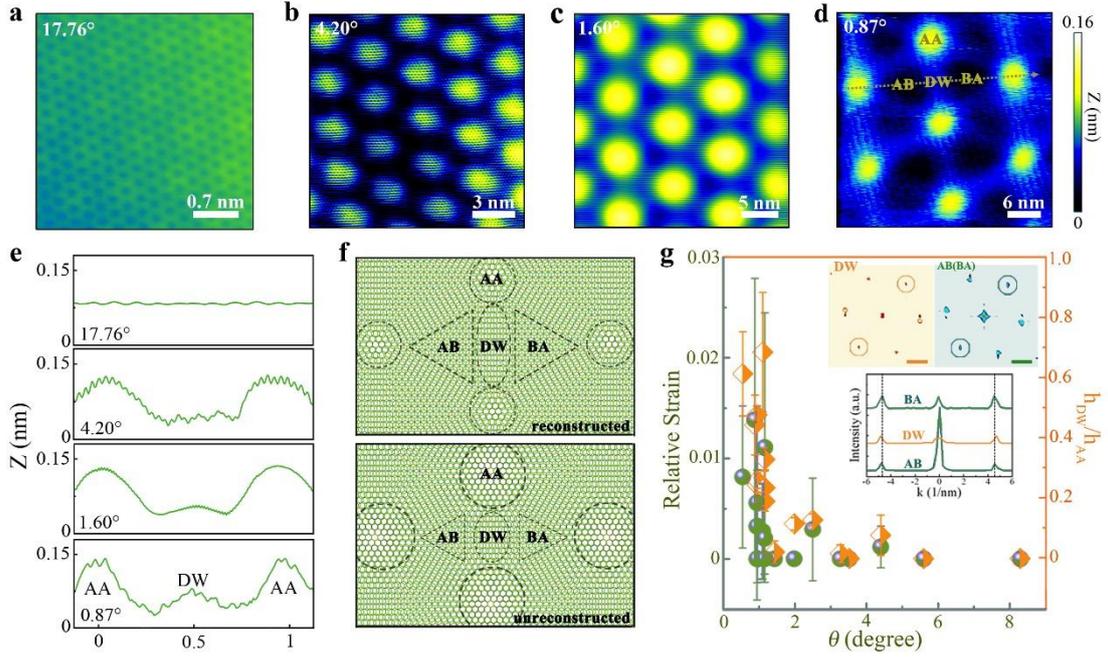

**Fig. 1. Characterizations of the structural transition with different twist angles. a-d** Typical STM images of TBG with $\theta = 17.76°$, $\theta = 4.20°$, $\theta = 1.60°$ and $\theta = 0.87°$, respectively ($V_s = 400$ mV, $I = 0.2$ nA). The AA, AB, BA and DW regions are marked in panel **d**. **e** Height profiles recorded at TBG with different twist angles along the yellow arrow, as shown in panel **d**. The x axis is normalized by the spacing between two AA stacking regions of different samples. **f** Schematic reconstructed (top panel) and unreconstructed (bottom panel) structures of the TBG in real space. **g** Relative strain and the height ratio of the TBG from 0.54° to 8.3°. Here, the relative strain is calculated according to ratio between lattice constant in the DW region and that in the AB/BA region. The average lattice constant in different regions is obtained according to the FFT in three directions. Height ratio is the relative height of the DW region to the AA region, as shown in panel **e**. Inset: representative FFT of the DW (top left) and AB/BA regions (top right). The six bright spots are reciprocal lattices of graphene. Scale bar: 2.5 nm$^{-1}$. Bottom panel: Section views cross the colored circles indicated in top panels. The dashed lines label the peaks obtained in the AB and BA regions.

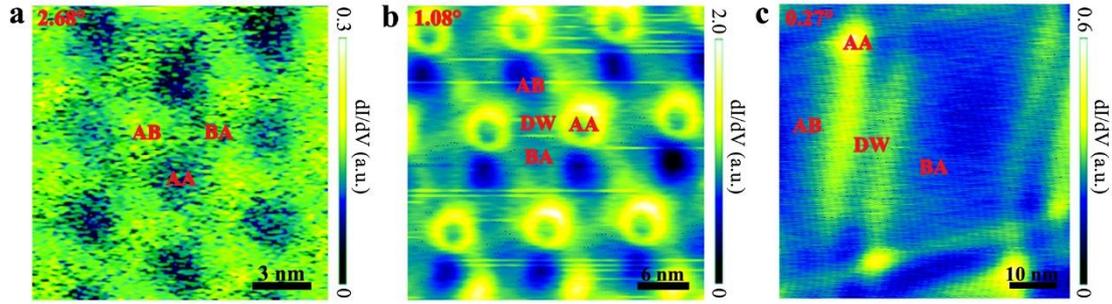

**Fig. 2. STS maps of TBG with different twist angles. a-c** *dI/dV* maps of the TBG with $\theta = 2.68°$, $\theta = 1.08°$ and $\theta = 0.27°$. The experimental parameters are **a** $V_s = -35$ mV; **b** $V_s = 20$ mV; **c** $V_s = 40$ mV. The AA, AB, BA and DW regions are marked in the maps.

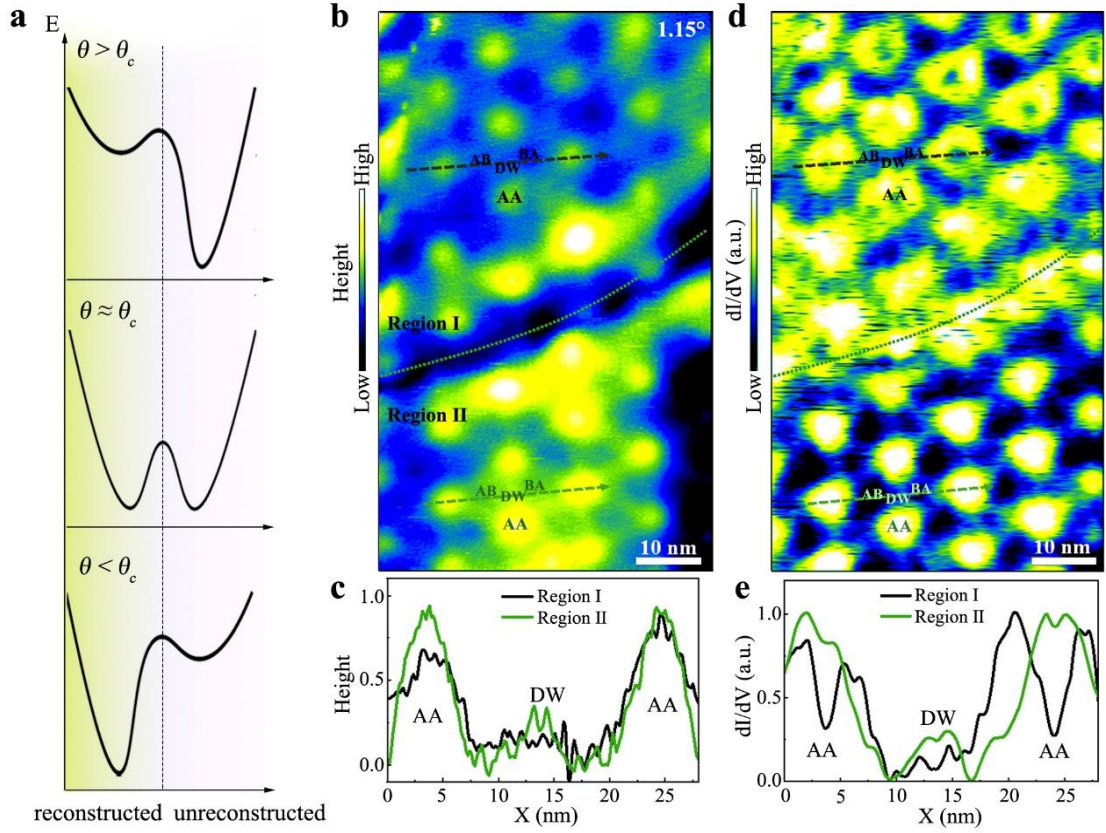

**Fig. 3. Bi-stable structures of a continuous TBG around the magic angle. a** The schematic energy profiles in the configuration space that illustrate the structural transition from unreconstructed TBG to reconstructed TBG as the twist angle decreases across the magic angle (or the $\theta_c$). **b** STM topographic image of a TBG showing both the unreconstructed and reconstructed structures ($V_S = 30$ mV, $I = 0.4$ nA). Region I (unreconstructed) and Region II (reconstructed) are separated by a dashed line. **c** Height profiles along dashed arrows in panel **b**. **d** Typical *dI/dV* map of the TBG taken at the same position as panel **b** at the energy of 30 meV. **e** *dI/dV* profiles along dashed arrows in panel **d**.

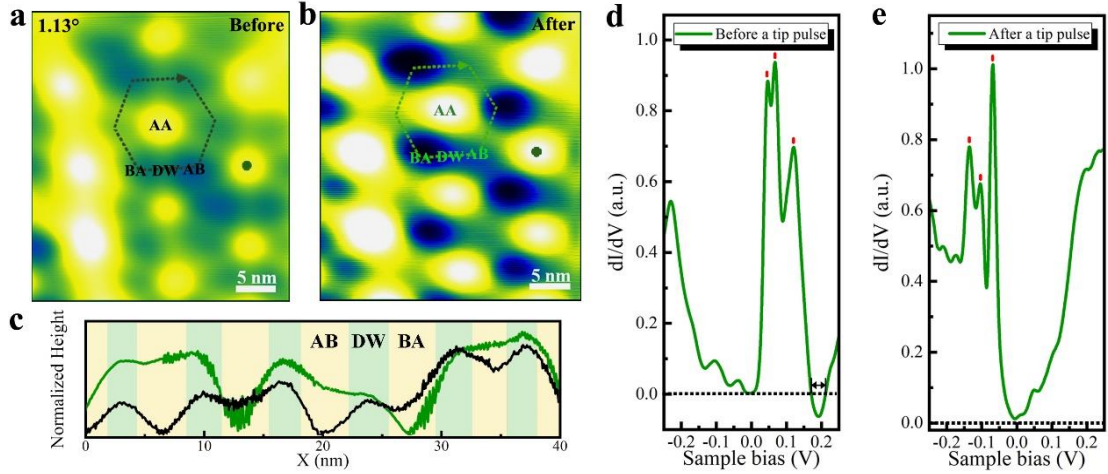

**Fig. 4. Tunable structures and electronic properties of the TBG around the magic angle. a** STM image ($V_S$ = 650 mV and $I$ = 0.3 nA) of a 1.13 °TBG showing the reconstructed structure. **b** STM image of the same region as panel **a** ($V_S$ = 650 mV and $I$ = 0.3 nA) after applying a 3-V tip pulse for 0.1 s. The STM tip is settled about 1 nm above the TBG. **c** Height profiles along dashed lines around a AA region in panels **a** and **b**. The DW and AB/BA regions are marked with different colors. **d, e** Typical STS spectra taken at the same position marked by solid dots in panels **a** and **b** respectively.